\documentclass[a4paper]{article}

\usepackage[english]{babel}
\usepackage[utf8]{inputenc}
\usepackage{ttpaper}
\usepackage{amssymb, amsthm, amsmath}
\usepackage{mathtools}
\usepackage{pgf,tikz}

\tikzstyle{vertex}=[circle, draw, inner sep=1pt, minimum size=1.5pt]
\usetikzlibrary{decorations.markings}
\newcommand{\vertex}{\node[vertex]}

\title{Smallest graphs achieving the Stinson bound}

\author{ M\'at\'e Gyarmati\inst1, P\'eter Ligeti\inst2}

\institute{\inst1 
         E\"otv\"os Lor\'and University\\
         Budapest, Martonv\'as\'ar, Hungary
         \\ e-mail: gyarmati93mate@gmail.com \\
         ORCID ID: 0000-0002-4122-9181 \\
         \inst2
         E\"otv\"os Lor\'and University\\
         Budapest, Martonv\'as\'ar, Hungary
         \\ e-mail:  turul@cs.elte.hu \\
         ORCID ID:  0000-0002-3998-0515\\
         }

\date{}

\begin{document}
\maketitle

\begin{abstract}
Perfect secret sharing scheme is a method of distribute a secret information $s$ among participants such that only predefined coalitions, called qualified subsets of the participants can recover the secret, whereas any other coalitions, the unqualified subsets cannot determine anything about the secret. The most important property is the efficiency of the system, which is measured by the information ratio. It can be shown that for graphs the information ratio is at most $(\delta+1)/2$ where $\delta$ is the maximal degree of the graph. Blundo et al. constructed  a family of $\delta$-regular graphs  with information ratio $(\delta+1)/2$ on at least $c\cdot 6^\delta$ vertices. We improve this result by constructing a significantly smaller graph family on $c\cdot 2^\delta$ vertices achieving the same upper bound both in the worst and the average case. 

\bigskip

\end{abstract}

\section{Introduction}
Secret sharing schemes were first introduced by Shamir \cite{shamir} and Blakley \cite{blakley} as a method of distribute a secret information $s$ among participants such that only predefined coalitions, called qualified subsets of the participants can recover the secret. If the unqualified subsets have no information about $s$, then we call it a perfect secret sharing. The set of the qualified subsets is called access structure. If all minimal elements of the access structure have exactly  two elements, then the access structure can be represented with a graph: let the participants denote the vertices, and two vertices are supposed to be connected by an edge if the respective participants are qualified together. 

One of the most frequently examined problem on secret sharing is the efficiency of a particular system. This can be characterized by the amount of information a given participant must remember correlate to the size of the secret. This amount is called the information ratio of the share of this participant. From the whole system point of view, it is possible to investigate two, slightly different quantities: the worst case information ratio is the information that the most heavily loaded participant must remember; or the average case, which is the average of the information ratios of the participants. Determining the information ratio is a challenging but interesting problem for both  cases even for small access structures.
 
The exact value of the worst-case information ratio was determined for most of the graphs with at most six vertices \cite{jacksonfive},\cite{blundo1},\cite{dijk6},\cite{sun_six}, \cite{finished_six} for trees \cite{csirmaztardos}, for $d$-dimensional cubes \cite{csirmaz_cube}, for graphs with large girth \cite{csl_girth}. On the other hand,  there are some sporadic results on the average case as well, see \cite{blundo_avr}, \cite{blundo1}, \cite{fulu_tree}, \cite{fulu_unicycle}. In  general however, there is a large gap between the  best known upper and lower bound. One notable universal upper bound was proved by Stinson in \cite{stinston_decconstr}. He showed that if the maximal degree of a graph is $\delta$, then the information ratios are at most $(\delta+1)/2$. Blundo et al.  in \cite{blundo_tightbounds} showed that this bound is sharp for an infinite graph class with maximal degree $\delta$. 

The graph class constructed by Blundo et al. with maximal degree $\delta$ has $2np^{\delta-3}$ vertices, where $n,p\ge6$. We give a significant improvement  of that result by presenting graph classes with much less vertices, i.e. $2^{\delta}$ for the worst case, and $3\cdot 2^{\delta-1}$ for the average case.



\section{Preliminaries}
Let $P$ be a finite set of participants. The access structure $\cal{A}$ on $P$ is a  subset of $2^P$ which is monotone increasing in the sense, that if $A\in \mathcal{A}$ and $A\subseteq B$, then $B\in \mathcal{A}$. 

\begin{definition} A perfect secret sharing scheme $S$ realizing $\mathcal{A}$ is a collection of random
variables $\xi_i$ for every $i\in \mathcal{P}$ and $\xi_s$ with a joint
distribution such that
\begin{itemize}
\item[(i)]  if $A\in \mathcal{A}$, then $\{\xi_i: i\in A\}$ determines $\xi_s;$
 \item[(ii)] if $A\notin \cal{A}$, then $\{\xi_i: i\in A\}$ is independent
of $\xi_s.$
\end{itemize}\end{definition}

Note that as a consequence of the monotonicity, the minimal elements of $\mathcal{A}$ determine $\mathcal{A}$. In this paper we consider only graph-based secret sharing schemes, where all the minimal elements of the access structure have two elements, hence from now on we will use a graph $G=(V,E)$ instead of an $\mathcal{A}.$

The size of a discrete random variable $\xi$ is measured by it's Shannon-entropy, defined as $H(\xi)=-\sum_j p_j\log(p_j)$, where $\xi$ has possible values $x_j$ with probabilities $p_j=P(\xi=x_j)$. The information ratio of a participant $v$ in $V$ is $H(\xi_v)/H(\xi_s)$.

\begin{definition}
Let $G=(V,E)$ be a graph. Then the information ratio of $G$ is
\begin{itemize}
\item $\sigma(G)=\inf\limits_S\max\limits_{v\in V} H(\xi_v)/H(\xi_s)$ in the worst case;
\item $\overline{\sigma}(G)=\inf\limits_S\ \sum\limits_{v\in V} \frac{H(\xi_v)}{H(\xi_s)|V|}$ in the average case
\end{itemize} where every infimum is taken over all perfect secret sharing schemes $\cal{S}$ realizing $G.$
\end{definition}

Every construction yields an upper bound for the information ratio, one notable example is the decomposition theorem of Stinson \cite{stinston_decconstr} yielding an upper bound derived from any covering of the graph. Let us mention, that this fundamental result has  more general possible interpretations though, here we present a simple consequence  of covering with stars only:

\begin{theorem}[Stinson-bound]\label{thm_stinson}
Let $G$ be a graph with maximal degree $\delta$. Then 
\begin{align}
\sigma(G)\leq (\delta+1)/2 \\
\overline{\sigma}(G)\leq (\delta+1)/2
\end{align}
\end{theorem}

Note that in the average case this bound can be sharp only for regular graphs.

Let $S$ be a perfect secret sharing scheme based on graph $G=(V,E)$ with shares $\xi_v$ for $v\in V$ and secret $\xi_s$, and define the values
$$f(A)=\frac{H(\xi_v:v\in A)}{H(\xi_s)}$$
for each $A\subseteq V$. Using the standard properties of the entropy function and the definition of the perfect secret sharing we get the following, so called Shannon-inequalities:
\begin{itemize}
\item[(a)] $f(\emptyset)=0$, and in general $f(A)\ge 0$ (positivity);
\item[(b)] if $A\subseteq B \subseteq V$ then $f(A)\le f(B)$ (monotonicity);
\item[(c)] $f(A) + f(B) \ge f(A\cap B) + f(A\cup B)$ (submodularity).
\item[(d)] if $A\subseteq B$, $A$ is an independent set and $B$ is not, then
   $f(A) + 1 \le f(B)$ (strong monotonicity);
\item[(e)] if neither $A$ nor $B$ is independent but $A\cap B$ is so, then
  $f(A)+f(B) \ge 1+f(A\cap B)+f(A\cup B)$ (strong submodularity).
\end{itemize}
Let $f$ be any function satisfying these five type of inequalities. Then $\max_{v\in V} f(v)$ is a lower bound for the information ratio in the worst case, and $\sum\limits_{v\in V} \frac{f(v)}{|V|}$ is a lower bound for the information ratio in the average case. This is the so-called entropy method. In some cases we shall use the rearrangement of the inequalities. 

For  simplicity we usually write $AB$ instead of $A\cup B$ for subsets of vertices, and $a$ instead of $\{a\}$ for vertices. 


\section{Results}

As we noted above, the information ratio of any graph is at most $(d+1)/2$, where $d$ is the maximal degree of the graph as a simple consequence of Stinson's decomposition theorem \cite{stinston_decconstr}. Within this section we construct infinite classes of graphs with this best achievable information ratio both in the worst and the average case. Additionally, these graphs are significantly smaller than the recently known examples satisfying this property. 

\subsection{Construction for the worst case}

Let $C_{d}^*$ be a graph with $2^{d+1}$ vertices, built from a $d$-dimensional cube, $2^d$ independent vertices and a 1-factor between the vertices of the cube and the independent vertex set.


Let $f$ be any function satisfying the Shannon-inequalities for $C_d^*$, and $A,B,X\subset V$ such that $A\subset X$. We shall use the following notation.



\begin{definition}
\begin{equation}
[[X,B,A]]=\sum_{b\in B}f(bX)-\sum_{a\in A}f(X-a) 
\end{equation}
\end{definition}

Split the vertices of $C_d^*$ into two equal independent sets the chessboard like fashion: $C_d^*=X_d\cup Y_d$, $|X_d|=|Y_d|=2^d$, $X_d\cap Y_d=\emptyset$, and $X_d$ and $Y_d$ are independent. 
Let $C_d$ be the $d$-dimensional subcube of $C_d^*$ and $A_d=C_d\cap X_d$ and $B_d=C_d\cap Y_d$. Clearly $A_d$ and $B_d$ are equal, independent and disjoint sets and $A_d\cup B_d=C_d$.

\begin{lemma}       \label{lem.1}
\begin{equation}
\sum_{v\in C_d} f(v)\geq [[X_d,B_d,A_d]]+d\cdot 2^{d-1} \label{l.egy}
\end{equation}
\end{lemma}

\proof First we check for $d=1$. $C_d^*$ is a path of length 3, with vertices $x,b,a,y$ respectively. The inequality in \ref{l.egy} becomes 
$$f(a)+f(b)\ge [[ax,b,a]]+1=f(abx)-f(x)+1.$$
Using the strong submodularity 
$$f(ab)+f(ax)\geq f(a)+f(abx)+1$$ and the submodularities 
\begin{align*}
f(a)+f(b)\geq f(ab) \\
f(a)+f(x)\geq f(ax),
\end{align*}
 we get:
$$f(abx)-f(x)\leq f(ab)+f(ax)-f(a)-1-f(x)\leq f(a)+f(b)-1,$$
which is the statement of the lemma.

Now suppose that \ref{l.egy} holds for the $d$ dimensional case. $C_{d+1}^*$ consist of two disjoint copy of $C_d^*$ and the vertices of the $d$-dimensional cubes are connected with a perfect matching $M$. Split this two copies of $C_d^*$ the chess-board like fashion to $X_d\cup Y_d$, and $X_d'\cup Y_d'$ respectively such that edges of $M$ are between $X_d$ and $Y_d'$, and $X_d'$ and $Y_d$. $A_d$, $B_d$, $A_d'$, $B_d'$, $A_{d+1}$, and $B_{d+1}$ are define as above. Clearly $A_d\cup A_d'=A_{d+1}$, $B_d\cup B_d'=B_{d+1}$, $X_d\cup X_d'=X_{d+1}$ and $Y_d\cup Y_d'=Y_{d+1}$, and $M$ is between $A_{d+1}$ and $B_{d+1}$. 

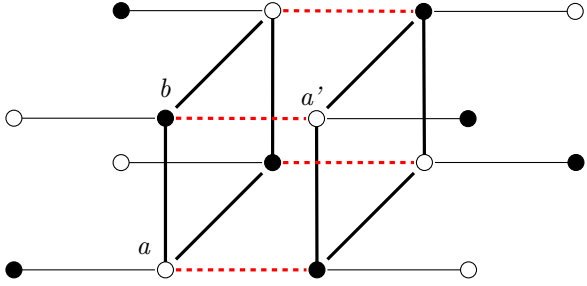
\begin{center}
		\begin{tikzpicture} [scale=2]
		\draw [line width= 1.2] (0:0) node (A){} -- (90:1) node (B){};
		\draw [line width= 1.2] (45:1) node (D){} -- (67.5:1.85) node (C){};
		\draw [line width= 1.2] (A) -- (D);
		\draw [line width= 1.2] (B) -- (C);
		\draw [line width= 1.2] (0:1) node (E){} -- (45:1.41) node (F){};
		\draw [line width= 1.2] (45:2.41) node (G){} -- (22.5:1.85) node (H){};
		\draw [line width= 1.2] (E) -- (H);
		\draw [line width= 1.2] (F) -- (G);
		\draw (A) -- (180:1) node (A1){};
		\draw (B) -- +(180:1) node (B1){};
		\draw (C) -- +(180:1) node (C1){};
		\draw (D) -- +(180:1) node (D1){};
		\draw (E) -- +(0:1) node (E1){};
		\draw (F) -- +(0:1) node (F1){};
		\draw (G) -- +(0:1) node (G1){};
		\draw (H) -- +(0:1) node (H1){};
		
		\draw [line width= 1.2, color=red, dash pattern=on 2pt off 2pt] (A) -- (E);
		\draw [line width= 1.2, color=red, dash pattern=on 2pt off 2pt] (B) -- (F);
		\draw [line width= 1.2, color=red, dash pattern=on 2pt off 2pt] (C) -- (G);
		\draw [line width= 1.2, color=red, dash pattern=on 2pt off 2pt] (D) -- (H);
		\draw [fill=white] (0:0) circle (1.5 pt);
		\draw [fill=black] (90:1) circle (1.5 pt);
		\draw [fill=white] (67.5:1.85) circle (1.5 pt);
		\draw [fill=black] (45:1) circle (1.5 pt);
		\draw [fill=black] (0:1) circle (1.5 pt);
		\draw [fill=white] (45:1.41) circle (1.5 pt);
		\draw [fill=black] (45:2.41) circle (1.5 pt);
		\draw [fill=white] (22.5:1.85) circle (1.5 pt);
		
		\draw [fill=black] (0:0)+(180:1) circle (1.5 pt);
		\draw [fill=white] (90:1)+(180:1) circle (1.5 pt);
		\draw [fill=black] (67.5:1.85)+(180:1) circle (1.5 pt);
		\draw [fill=white] (45:1)+(180:1) circle (1.5 pt);
		\draw [fill=white] (0:1)+(0:1) circle (1.5 pt);
		\draw [fill=black] (45:1.41)+(0:1) circle (1.5 pt);
		\draw [fill=white] (45:2.41)+(0:1) circle (1.5 pt);
		\draw [fill=black] (22.5:1.85)+(0:1) circle (1.5 pt);
		
		\draw (90:1.2) node (B2){\textit{b}};
		\draw (135:0.2) node (A2){\textit{a}};
		\draw (50:1.5) node (C2){\textit{a'}};
		\end{tikzpicture}
\end{center}

Let $b\in B_d$ arbitrary, and its pair in $M$ is $a'\in A_d'$. By submodularity 
\begin{align}
	f(bX_d)-f(X_d) \geq f(bX_dX_d'-a')-f(X_dX_d'-a') \label{ineq.0}
\end{align}

Let $a\in A_d$ be arbitrary neighbor of $b$. Then $bX_d'$ and $abX_d'-a'$ are both qualified, but their intersection, $bX_d'-a'$ is independent, hence by the strong submodularity. 
\begin{align}
	f(bX_d')-f(bX_d'-a')\geq f(abX_d')-f(abX_d'-a')+1 \label{ineq.1}
\end{align}

Using the submodularity, we can write two additional inequalities:
\begin{align}
	f(X_d)-f(X_d'-a')&\geq f(bX_d')-f(bX_d'-a') \label{ineq.2} \\
f(abX_d')-f(abX_d'-a') &\geq f(bX_dX_d')-f(bX_dX_d'-a') \label{ineq.3}	 
 \end{align}

Adding the inequalities \ref{ineq.0}, \ref{ineq.1}, \ref{ineq.2}, \ref{ineq.3}, we get
\begin{align}
	f(bX_d)-f(X_d)+f(X_d')-f(X_d'-a') \geq f(bX_dX_d')-f(X_dX_d'-a')+1 \label{ineq.01}
\end{align}

Similarly if we choose arbitrary $b'\in B_d'$ and $a\in A_d$ is the pair of $b'$ in $M$, then
\begin{align}
	f(b'X_d')-f(X_d')+f(X_d)-f(X_d-a) \geq f(b'X_dX_d')-f(X_dX_d'-a)+1 \label{ineq.02}
\end{align}

All edges between $B_d$ and $A_d'$, and $B_d'$ and $A_d$ yield an inequality as \ref{ineq.01} and \ref{ineq.02} respectively. Adding up all these $2^d$ inequalities, on the left hand side all $f(X_d)$ and $f(X_d')$ cancel out the remaining is
\begin{align*}
\sum_{b\in B_d} f(bX_d)-\sum_{a'\in A'_d} f(X_d'-a')+&\sum_{b'\in B'_d} f(b'X_d')-\sum_{a\in A_d} f(X_d-a)
\ge \\ 
\sum_{b\in B_d} f(bX_{d}X_d')-\sum_{a'\in A'_d} f(X_{d}X_d'-a')+&\sum_{b'\in B'_d} f(b'X_{d}X_d')-\sum_{a\in A_d} f(X_{d}X_d'-a)+2^d.
\end{align*}

Using the facts that $X_d \cup X_d'=X_{d+1}$, $A_d\cup A_d'=A_{d+1}$, and $B_d\cup B_d'=B_{d+1}$ this can be written as
\begin{align}
[[X_d,B_d,A_d]]+[[X_d',B_d',A_d']]\geq[[X_{d+1},B_{d+1},A_{d+1}]]+2^d. \label{ineq.11}
\end{align}

Applying the inductive hypothesis we get 

\begin{align*}
	[[X_d,B_d,A_d]]+[[X_d',B_d',A_d']] &\leq \\  
    \sum_{v\in C_d} f(v)-d\cdot 2^{d-1}  +\sum_{v\in C_d'} f(v) -d\cdot 2^{d-1}  &=\sum_{v\in C_{d+1}} f(v) -d\cdot 2^d 
    \end{align*}

This inequality and  \ref{ineq.11} together yield the statement of the lemma. \qed

\begin{lemma} \label{lem.2}
\begin{align}
[[X_d,B_d.A_d]]\geq 2^d
\end{align}
\end{lemma}
\proof
$C_d$ is a regular bipartite graph with parts $A_d$ and $B_d$ hence there is a perfect matching between $A_d$ and $B_d$. Let $(a,b)$ be a pair in the perfect matching, $a\in A_d$, $b\in B_d$, and let $y$ be the leaf neighbor of $a$. $X_d$ and $X_d-a$ are independent, but $bX_d$ and $yX_d$ are not, thus by the strong monotonicity and the submodularity:
\begin{align*}
f(bX_d)-f(X_d)\geq 1 \\
f(yX_d)-f(yX_d-a)\geq 1 \\
f(yX_d-a)+f(X_d)-f(X_d-a)-f(yX_d) \geq 0
\end{align*}

The sum of this three inequality gives
\begin{align}
f(bX_d)-f(X_d-a)\geq 2 \label{ineq.lem2}
\end{align}

 Adding up \ref{ineq.lem2} for all the $2^{d-1}$ edges in the perfect matching between $A_d$ and $B_d$ gives the statement of the lemma \qed

\begin{lemma} \label{lem.3}
$$\sum_{v\in C_d} f(v)\geq (d+2)2^{d-1}$$
\end{lemma}

\proof
This is just an easy corollary of lemma \ref{lem.1} and lemma \ref{lem.2}.
\qed

\begin{theorem} \label{worst_case}
 The information ratio of $C_{\delta-1}^*$ in the worst case is 
$$\sigma(C_{\delta-1}^*)=\frac{\delta+1}{2}$$
\end{theorem}

\proof
The maximal degree of $C_{\delta-1}^*$ is $\delta$, hence we get $\sigma(C_{\delta-1}^*)\le\frac{\delta+1}{2}$ from the Stinson-bound \ref{thm_stinson}.

On the other hand, $C_{\delta-1}$ has $2^{\delta-1}$ vertices, hence by Lemma \ref{lem.3} for at least one vertex $$f(v)\geq \frac{(\delta+1)2^{\delta-2}}{2^{\delta-1}}=\frac{\delta+1}{2}$$ 
holds, which completes the proof. \qed

\subsection{Construction for the average case}

We construct the graph class $\Delta_d$ as follows. Let $C_d^{(0)}$, $C_d^{(1)}$ and $C_d^{(2)}$ be three disjoint $d$-dimensional cubes. The vertices of each cube can divide into two independent sets of vertices the chessboard-like fashion, $C_d^{(i)}=A_d^{(i)}\cup B_d^{(i)}$. Now we get $D_d$ by adding arbitrary 1-factors between $A_d^{(i)}$ and $B_d^{(i+1)}$ for $i=1,2,3$. The family $\Delta_d$ consists of all graphs $D_d$ constructed by this method, see the following example of a $D_3\in\Delta_3:$

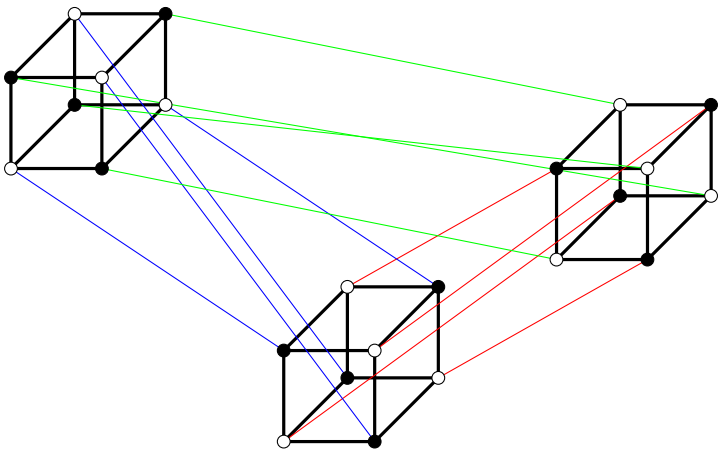
\begin{center}
\begin{tikzpicture}[scale=1.2]
		\vertex at (0,0) (A){};
		\vertex at (1,0) (B){};
		\vertex at (1,1) (C){};
		\vertex at (0,1) (D){};
		\draw [line width= 1.2] (A) -- (B);
		\draw [line width= 1.2] (C) -- (D);
		\draw [line width= 1.2] (A) -- (D);
		\draw [line width= 1.2] (B) -- (C);
		\vertex at (0.7,0.7) (E){};
		\vertex at (1.7,0.7) (F){};
		\vertex at (1.7,1.7) (G){};
		\vertex at (0.7,1.7) (H){};		
		\draw [line width= 1.2] (E) -- (F);
		\draw [line width= 1.2] (H) -- (G);
		\draw [line width= 1.2] (E) -- (H);
		\draw [line width= 1.2] (F) -- (G);
		\draw [line width= 1.2] (A) -- (E);
		\draw [line width= 1.2] (B) -- (F);
		\draw [line width= 1.2] (C) -- (G);
		\draw [line width= 1.2] (D) -- (H);

		\vertex at (3,2) (A2){};
		\vertex at (4,2) (B2){};
		\vertex at (4,3) (C2){};
		\vertex at (3,3) (D2){};
		\draw [line width= 1.2] (A2) -- (B2);
		\draw [line width= 1.2] (D2) -- (C2);
		\draw [line width= 1.2] (A2) -- (D2);
		\draw [line width= 1.2] (B2) -- (C2);
		\vertex at (3.7,2.7) (E2){};
		\vertex at (4.7,2.7) (F2){};
		\vertex at (4.7,3.7) (G2){};
		\vertex at (3.7,3.7) (H2){}; 
		\draw [line width= 1.2] (E2) -- (F2);
		\draw [line width= 1.2] (H2) -- (G2);
		\draw [line width= 1.2] (E2) -- (H2);
		\draw [line width= 1.2] (F2) -- (G2);
		
		\draw [line width= 1.2] (A2) -- (E2);
		\draw [line width= 1.2] (B2) -- (F2);
		\draw [line width= 1.2] (C2) -- (G2);
		\draw [line width= 1.2] (D2) -- (H2);
		
		\vertex at (-3,3) (A3){};
		\vertex at (-2,3) (B3){};
		\vertex at (-2,4) (C3){};
		\vertex at (-3,4) (D3){};
		\draw [line width= 1.2] (A3) -- (B3);
		\draw [line width= 1.2] (D3) -- (C3);
		\draw [line width= 1.2] (A3) -- (D3);
		\draw [line width= 1.2] (B3) -- (C3);
		
		\vertex at (-2.3,3.7) (E3){};
		\vertex at (-1.3,3.7) (F3){};
		\vertex at (-1.3,4.7) (G3){};
		\vertex at (-2.3,4.7) (H3){};
		\draw [line width= 1.2] (E3) -- (F3);
		\draw [line width= 1.2] (H3) -- (G3);
		\draw [line width= 1.2] (E3) -- (H3);
		\draw [line width= 1.2] (F3) -- (G3);
		
		\draw [line width= 1.2] (A3) -- (E3);
		\draw [line width= 1.2] (B3) -- (F3);
		\draw [line width= 1.2] (C3) -- (G3);
		\draw [line width= 1.2] (D3) -- (H3);

		\draw [color=red](H) -- (D2);
		\draw [color=red](F) -- (B2);
		\draw [color=red](A) -- (E2);
		\draw [color=red](C) -- (G2);
		
		\draw [color=blue](A3) -- (D);
		\draw [color=blue](C3) -- (B);
		\draw [color=blue](F3) -- (G);
		\draw [color=blue](H3) -- (E);
		
		\draw [color=green](A2) -- (B3);
		\draw [color=green](C2) -- (E3);
		\draw [color=green](F2) -- (D3);
		\draw [color=green](H2) -- (G3);
		
		\draw [fill=white] (0,0) circle (2 pt);
		\draw [fill=black] (1,0) circle (2 pt);
		\draw [fill=white] (1,1) circle (2 pt);
		\draw [fill=black] (0,1) circle (2 pt);
		\draw [fill=black] (0.7,0.7) circle (2 pt);
		\draw [fill=white] (1.7,0.7) circle (2 pt);
		\draw [fill=black] (1.7,1.7) circle (2 pt);
		\draw [fill=white] (0.7,1.7) circle (2 pt);
		
		\draw [fill=white] (3,2) circle (2 pt);
		\draw [fill=black] (4,2) circle (2 pt);
		\draw [fill=white] (4,3) circle (2 pt);
		\draw [fill=black] (3,3) circle (2 pt);
		\draw [fill=black] (3.7,2.7) circle (2 pt);
		\draw [fill=white] (4.7,2.7) circle (2 pt);
		\draw [fill=black] (4.7,3.7) circle (2 pt);
		\draw [fill=white] (3.7,3.7) circle (2 pt);		

		\draw [fill=white] (-3,3) circle (2 pt);
		\draw [fill=black] (-2,3) circle (2 pt);
		\draw [fill=white] (-2,4) circle (2 pt);
		\draw [fill=black] (-3,4) circle (2 pt);
		\draw [fill=black] (-2.3,3.7) circle (2 pt);
		\draw [fill=white] (-1.3,3.7) circle (2 pt);
		\draw [fill=black] (-1.3,4.7) circle (2 pt);
		\draw [fill=white] (-2.3,4.7) circle (2 pt);		
\end{tikzpicture}
\end{center}


\begin{theorem} \label{average_case}
The average information ratio of every  graph  $D_{\delta-1}\in\Delta_{\delta-1}$ is
$$\overline{\sigma}(D_{\delta-1})=\frac{\delta+1}{2}$$
\end{theorem}

\proof All vertices of any graphs $D_{\delta-1}$ have a degree of $\delta$, hence by Stinson-bound \ref{thm_stinson} the information ratio is at most $(\delta+1)/2$.

Restrict an arbitrary fixed graph $D_{\delta-1}\in\Delta_{\delta-1}$ to $H=C_{\delta-1}^{(1)}\cup B_{\delta-1}^{(2)}\cup A_{\delta-1}^{(3)}$. It is easy to see, that $H$ is isomorphic to $C_{\delta-1}^*$. $C_{\delta-1}^{(1)}$ is a $d-1$ dimensional cube, and $B_{\delta-1}^{(2)}\cup A_{\delta-1}^{(3)}$ is an independent set. Furthermore there is a perfect matching between $A_{\delta-1}^{(1)}$and $B_{\delta-1}^{(2)}$, and between $B_{\delta-1}^{(1)}$ and $A_{\delta-1}^{(3)}$, hence there is a perfect matching between $C_{\delta-1}^{(1)}$ and $B_{\delta-1}^{(2)}\cup A_{\delta-1}^{(3)}$. This shows that $H$ is an induced subgraph of $D_{{\delta-1}}$ isomorphic to $C_{\delta-1}^*$ hence we can apply lemma \ref{lem.3} for $H$.

\begin{align}
\sum_{v\in C_{\delta-1}^{(1)}} f(v)\geq (\delta+1)2^{\delta-2} \label{aver.ineq1}
\end{align}

Similarly, if we consider $C_{\delta-1}^{(2)}\cup B_{\delta-1}^{(3)}\cup A_{\delta-1}^{(1)}$ and $C_{\delta-1}^{(3)}\cup B_{\delta-1}^{(1)}\cup A_{\delta-1}^{(2)}$ we obtain
\begin{align}
\sum\limits_{v\in C_{\delta-1}^{(2)}} f(v) \geq (\delta+1)2^{\delta-2} \label{aver.ineq2} \\
\sum\limits_{v\in C_{\delta-1}^{(3)}} f(v) \geq (\delta+1)2^{\delta-2} \label{aver.ineq3}
\end{align}
Combining \ref{aver.ineq1}, \ref{aver.ineq2}, \ref{aver.ineq3}, we get 
\begin{equation}
\sum_{v\in D_{\delta-1}} f(v)\geq 3(\delta+1)2^{\delta-2}.
\end{equation}
$D_{\delta-1}$ has $3\cdot 2^{\delta-1}$ vertices, hence the information ratio in the average case is at least $(\delta+1)/2$, which completes the proof. \qed


\section{Conclusion}

In this paper we present new families of graphs of maximal degree $\delta$ achieving the best possible information ratio value given by the Stinson bound $(\delta+1)/2$. These graphs are asymptotically the smallest ones recently: the first graph class achieves the bound in the worst case on $2^{\delta}$ vertices and the other graph class constructed for the average case has $3/2\cdot2^{\delta}$ vertices in contrast to the best known constructions on $c\cdot6^\delta$ vertices.

\section*{Acknowledgement}

This research has been partially supported by  the European Union, co-financed by the European Social Fund (EFOP-3.6.2-16-2017-00013, Thematic Fundamental Research Collaborations Grounding Innovation in Informatics and Infocommunications). The authors thank the members of the Crypto Group of the R\'enyi Institute, and especially L\'aszl\'o Csirmaz and G\'abor Tardos, the friutful comments and discussions.


\begin{thebibliography}{9}


\bibitem{blakley}
G.~R. Blakley, Safeguarding cryptographic keys, in: \emph{Proc. of the Nat.
  Comp. Conf.}, \textbf{48}, (1979) pp.~313--317.

\bibitem{blundo_tightbounds} C. Blundo, A. De Santis, R. De Simone, U. Vaccaro, Tight bounds on the information rate of
secret sharing schemes, Des., Codes and Crypt. \textbf{11} (2) (1997) pp. 107--110.

\bibitem{blundo_avr} C. Blundo, A. De Santis, L. Gargano, U. Vaccaro, On the information rate of
secret sharing schemes, Theor. Comp. Sci. \textbf{154} (2) (1996) pp. 283--306.

\bibitem{blundo1}
C.~Blundo, A.~De Santis, D.~R. Stinson and U.~Vaccaro, Graph decomposition and
  secret sharing schemes, \emph{J. of Crypt.} \textbf{8} (1995), 39--64.
  
\bibitem{stinson6}
E.~F. Brickell and D.~R. Stinson, Some improved bounds on the information rate
  of perfect secret sharing schemes, \emph{J. of Crypt.} \textbf{5} (1992),
  153--166.
   
\bibitem{csirmaz_log}
L.~Csirmaz, Secret sharing schemes on graphs, \emph{Studia Math. Hung.}
  \textbf{44} (2007), 297--306. 

\bibitem{csirmaz_cube}
Csirmaz, L.: Secret sharing on the d-dimensional cube. \emph{Des. Codes Cryptogr.} 74, 719–729
(2015)

\bibitem{csl_girth}
L.~Csirmaz and P.~Ligeti, Secret sharing on large girth graphs, submitted to  \emph{Crypt. and Comm.} (2018)

\bibitem{csirmaztardos}
L.~Csirmaz and G.~Tardos, Optimal information rate of secret sharing schemes on
  trees, \emph{IEEE Tr. on Inf. Theory} \textbf{59} (2013), 2527--2630.


\bibitem{dijk6}
M.~van Dijk, On the information rate of perfect secret sharing schemes,
  \emph{Des., Codes and Crypt.} \textbf{6} (1995), 143--160.  

\bibitem{finished_six}
O.~Farr\`as, T.~Kaced, S.~Martin, C.~Padr{\'o}, Improving the Linear Programming Technique in the Search for Lower Bounds in Secret Sharing. \emph{Cryptology ePrint Archive, Report} 2017/919,2017; available at https://eprint.iacr.org/2017/919

\bibitem{fulu_tree}
H.-L. Fu, H.-C. Lu, The exact values of the average information ratio for tree-based access structures of perfect secret sharing schemes, \emph{Des., Codes and Crypt.} \textbf{73} (1) (2014), 37--46. 

\bibitem{fulu_unicycle}
H.-L. Fu, H.-C. Lu, The optimal average information ratio of secret-sharing schemes for the access structures based on unicycle graphs and bipartite graphs, \emph{Disc. Appl. Math.} \textbf{233}  (2017), 131--142. 


\bibitem{jacksonfive}
W.~Jackson and K.~M. Martin, Perfect secret sharing schemes on five
  participants, \emph{Des., Codes and Crypt.} \textbf{9} (1996), 233--250.

\bibitem{shamir}
A.~Shamir, How to share a secret, \emph{Comm. of the ACM} \textbf{22} (1979),
  612--613.

\bibitem{stinston_decconstr} D. R. Stinson, Decomposition construction for secret sharing schemes, IEEE Trans. on Inf.
Theory, 40 (1) (1994) pp. 118–125.

\bibitem{sun_six}
H.L. Sun and B.L. Chen, Weighted decomposition construction for perfect secret
  sharing schemes, \emph{Comp. and Math. with Appl.} \textbf{43} (2002),
  877--887.



\end{thebibliography}
\end{document}